\documentstyle[prl,aps,twocolumn]{revtex}

\begin{document}

\draft
\title{ Semiclassical Theory of Short Periodic Orbits in Quantum Chaos}
\author{ Eduardo G. Vergini } 

\address{ Departamento de F\'{\i}sica, Comisi\'on Nacional de
Energ\'{\i}a At\'omica. 
 Av. del Libertador 8250, 1429 Buenos Aires, Argentina.}

\date{\today}
\maketitle

\begin{abstract}
We have developed a semiclassical theory of short periodic orbits to
obtain all quantum information of a bounded chaotic Hamiltonian system.
If $T_{1}$ is the period of the shortest periodic orbit, $T_{2}$ the
period of the next one and so on, the number $N_{p.o}$ of periodic 
orbits required in the calculation is such that 
$T_{1}+\cdots+T_{N_{p.o}}\simeq T_{H}$, with $T_{H}$ the Heisenberg
time. As a result $N_{p.o} \simeq h T_{H}/\ln (h T_{H})$, where $h$ is the
topological entropy. For methods related to the trace formula 
$N_{p.o} \simeq \exp(h T_{H})/ (h T_{H})$.
\end{abstract}

\centerline{PACS numbers: 05.45.+b, 03.65.Sq, 03.20.+i}

%
%

\narrowtext
The semiclassical evaluation of energy spectra in classically
chaotic Hamiltonian systems was started in 1971 with the 
Gutzwiller's trace
formula \cite{gut}. From then on a big effort was dedicated to 
extend the formalism to wave functions \cite{bog} and to use 
resummation techniques to improve convergence properties of trace 
formulas \cite{vor,ber}.
However, a common drawback in all those approaches is the
requirement of an enormous number of periodic orbits ({\bf P.O}s), restricting
explicit calculations to very special systems; e.g., systems where classical
mechanics is handled by a symbolic dynamics. Moreover, these approaches miss
a simple understanding of wave mechanics in terms of classical objects.  

Recently it was shown that chaotic eigenfunctions 
can be described in terms of
a small number of localized structures living on short periodic
orbits \cite{ver}. Having this 
in mind, we
present a simple semiclassical formalism to obtain quantum mechanics
from a very small number of short {\bf P.O}s. In order to do so,
we will construct resonances  (or scar functions) \cite{coment} of short {\bf P.O}s explicitly, evaluating the interaction between them.

Let $\gamma$ be an unstable {\bf P.O} isolated on each energy
surface. We introduce a curvilinear coordinate system choosing
the $x$ axis along the trajectory and the $y$ axis perpendicular to
it at $x$ (to simplify the exposition we take only one transversal
direction). On the {\bf P.O} $y\!=\!0$. Classical mechanics in a
neighborhood of the orbit is governed by a transversal
symplectic matrix $M(x)$ of elements $m_{ij}(x)$ ($i,j=1,2$), which
describes the linearized motion on the energy surface. Then, a
point with transversal coordinates ( $y$, $p_{y}$) at $x\!=\!0$ evolves
according to the following rule:
\[
\left( \begin{array}{c} 
		y(x) \\
		p_{y}(x)
	         \end{array} \right)
= \left( \begin{array}{cc}
	    m_{11}(x)   &   m_{12}(x)  \\
	    m_{21}(x)   &   m_{22}(x)
		\end{array}  \right)
\left( \begin{array}{c}
	       y   \\
	       p_{y}
		\end{array}   \right).
\]
We have selected the origin $x\!=\!0$ such that $m_{11}(L)=m_{22}(L)$,
with $L$ the length of $\gamma$ \cite{ac0}. There are at 
least $2\nu$ points on
the orbit satisfying that condition, with $\nu$ the Maslov index.
With this choice the monodromy matrix $M(L)$ acquires the form,
\[ 
 M(L)=(-1)^{\nu} \left( \begin{array}{cc}
                \cosh (\lambda L)  &  \sinh (\lambda L)/\tan (\varphi) \\
 \sinh (\lambda L)\; \tan(\varphi) &   \cosh (\lambda L)
                                 \end{array}  \right).
\]        
$\lambda$ is the Lyapunov exponent in units of $[length^{-1}]$ and
$\tan(\varphi)\;(\ne 0)$ in units of $[momentum/length]$ defines the
slope of the unstable manifold in the plane $y-p_{y}$ 
(the slope of the stable manifold being $-\tan (\varphi)$).

In general it is impossible to compare vectors living in the plane
$y-p_{y}$ because the axes have different units. However, when 
the directions are symmetrical with respect to the axes it is only
necessary to compare one component. Then, we change to new axes
$\xi_{u}$ and $\xi_{s}$, on the unstable and stable manifolds 
respectively, such that their 
projections on each axis are equal in absolute value. The 
symplectic matrix $B$ transforming coordinates from the new axes 
into the old ones is
\[ 
B=(\xi_{u}\;\xi_{s})= (1/ \sqrt{2}) \left( \begin{array}{cc}
			  	1/\alpha   &   -s/\alpha  \\
				s\; \alpha   &      \alpha 
				\end{array}  \right) ,
\]
with $\;\alpha=\sqrt{|\tan (\varphi )|}$ and $s={\rm sign} (\varphi)$.
Observe that 
$\;B^{-1}M(L)B=(-1)^{\nu} \exp (\lambda L D ) \;$,
with $D$ a diagonal matrix of elements $d_{11}=1$ and $d_{22}=-1$.

Now, we decompose $M(x)$ into a periodic matrix $F(x)$ describing
the evolution of the manifolds, and a matrix (depending in a simple
way of $x$) describing the exponential contraction-dilation along the
manifolds, 
\begin{equation}
M(x)=F(x)\; \exp (x \lambda K)\equiv
F(x)\; B \;\exp (x \lambda D)\; B^{-1},
\label{de}
\end{equation}
with $K \equiv BDB^{-1}$ [$k_{11}\!=\!k_{22}\!=\!0,\;k_{21}\!=\!
1/k_{12}\!=\!\tan (\varphi)$].
Equation (\ref{de})  defines 
$F(x)$ in terms of $M(x)$ and we can 
see that $F(L)=(-1)^{\nu}\openone$. Floquet's
theorem \cite{flo} affirms that the decomposition given in 
(\ref{de}) can also be obtained for systems with 
many transversal directions.

Now, it is possible to construct a family of resonances 
associated to $\gamma$. 
We adapt well known semiclassical techniques
to obtain eigenfunctions concentrated in the neighborhood of 
stable {\bf P.O}s \cite{bab}. In the
$x$ direction we consider the typical solution of a one dimensional
motion, and in the transversal
one we use a wave packet which evolves according to $F(x)$,
\begin{equation}
\psi_{\gamma}(x,y)=\frac{\exp \{i\;[S(x)+y^{2}\;\Gamma(x)/2]/
\hbar-i \;\phi(x)/2\} } {\sqrt{T \;\dot x}\;\;\; 
[\pi\;(\hbar/J)\; 
|Q(x)|^{2}]^{1/4}},
\label{res}
\end{equation}
where $S(x)\!=\!\int_{0}^{x}m{\dot x}\;dx$ is the action, $T$
the period of $\gamma$ and $J$ the unit area in the plane $y-p_{y}$. 
$\Gamma (x)\equiv P(x)/Q(x)$, with $Q(x)$ [$P(x)$] the $y$ ($p_{y}$)
component of the complex vector \cite{gausiana}
\begin{equation} 
\xi_{u}(x)+\!i \;\xi_{s}(x) \equiv F(x)(\xi_{u}+\!i \; 
\xi_{s})=M(x)B \left( \begin{array}{c}
		    \!\!  e^{-x \lambda} \!\! \\
		  \!\! i \; e^{ x \lambda} \!\!
			\end{array} \right). 
\label{vec}
\end{equation}
Equation (\ref{vec}) shows that it is not necessary to evaluate explicitly
$F(x)$.
The area preserving property of $F(x)$ guarantees the following
normalization condition:
\begin{equation}
Q^{*}(x)\;P(x)-Q(x)\;P^{*}(x)=2 i \; \xi_{u}(x) \wedge
\xi_{s}(x)=2i J.
\label{nor}
\end{equation}
Then, ${\rm Im} [\Gamma (x)]=J/|Q(x)|^{2}>0$. Accordingly 
$\psi_{\gamma}$ is concentrated around $\gamma$
and well-behaved in all regions except in the
neighborhood of a turning point ($\dot x=0$) \cite{ac2}. 
Finally, the complex number
$Q(x)$ sweeps an angle $\phi (x)$ while evolving from $0$ to $x$,
and  $\nu \equiv \phi(L)/\pi$ (this is the definition of the
Maslov index).

$\psi_{\gamma}$ is a continuous function at $x\!=\!L$ if the
accumulated phase around the orbit is an integral multiple of
$2\pi$. This condition determines the admitted energies 
$E_{\gamma}$ of the {\bf P.O} and
corresponds to the Bohr-Sommerfeld
quantization rule: $S(L)/\hbar-\nu \pi/2= 2n \pi$, where
$n=0,1,\ldots$ is the number of excitations
along $\gamma$.

We stress that the semiclassical construction of eigenfunctions
in the neighborhood of {\it stable} orbits is similar
to (\ref{res}). The initial complex vector of equation (\ref{vec})
is replaced by the eigenvector of the monodromy matrix (a complex
vector in this case) satisfying (\ref{nor}). And of course, the evolution
of the vector is given by the transversal symplectic matrix
without modifications. Eigenvalues have an error
${\cal O}(\hbar)$ and eigenfunctions
an error ${\cal O}(\sqrt{\hbar})$. Moreover,
it is possible to improve the accuracy by including transversal
excitations \cite{bab}. 

In our case, there is an essential error because the evolution in 
(\ref{vec}) is given by a modified transversal matrix.
In order to eliminate that error, we will first evaluate the action
of the semiclassical evolution operator for infinitesimal times
over the resonance in the form
\begin{equation}
\hat{H} \equiv i \hbar \; {\lim}_{\delta t \rightarrow 0} 
(\hat{U} (\delta t)-\hat{1} )/\delta t .
\label{hamil}
\end{equation}
The classical transversal evolution from $x$ to $x+\delta x$ is
given by $M_{x}(\delta x)\!=\!M(x+\delta x)\; M(x)^{-1}$. So,
$M_{x}(\delta x)\; F(x)\!=\!F(x+\delta x)\; \exp (\delta x\; \lambda K)$.
Observing that $K\xi_{u}\!=\!\xi_{u}$ and $K\xi_{s}\!=\!-\xi_{s}$,
it results to first order in $\delta x$:
\begin{equation}
\begin{array}{l}
M_{x}(\delta x)\;\xi_{u}(x)\simeq(1+\delta x\;\lambda)\;\xi_{u}(x+\delta x),\\
M_{x}(\delta x)\;\xi_{s}(x)\simeq(1-\delta x\;\lambda)\;\xi_{s}(x+\delta x).
\end{array}
\label{relations}
\end{equation}
The above expressions actually show clearly the approximation involved
in the construction. We have forced the vector $\xi_{u}$ ($\xi_{s}$) to 
evolve without dilation (contraction) while the right evolution dilates
(contracts) the vector with a rate 
specified by $\lambda$. Then, we set $\delta x \! = \! \dot{x} \;\delta t$
and use (\ref{relations}) to evaluate the action of $\hat{U}(\delta t)$
on $\psi_{\gamma}$. Finally, after some calculations
we obtain from (\ref{hamil}), 
\begin{equation}
\hat{H} \psi_{\gamma}(x,y)=g_{\gamma}(x,y)\;\psi_{\gamma}(x,y) ,
\label{int}
\end{equation}
with $\;g_{\gamma}(x,y)= E_{\gamma}\!+\! i 
\hbar \dot{x} \lambda \; ( y^{2}J/\hbar-
|Q(x)|^{2}/2 )/ Q(x)^{2}$.\\
\\
Then, the application of the semiclassical Hamiltonian operator to the
resonance gives the term $E_\gamma \psi_{\gamma}$ as expected, plus a
resonance of $\gamma$ with two excitations in the transversal direction.
In fact, the two excitations are also expected because 
the right evolution
produces a quadrupole-like deformation of the wave packet.
Equation (\ref{int}) is a extremely powerful 
tool: it is the key to evaluate matrix elements between
{\bf P.O}s.
As the operator $\hat{H}$ is not exactly Hermitian for finite values of 
$\hbar$, we define a symmetrized interaction
between two {\bf P.O}s $\gamma$ and $\delta$ as follows 
(in Dirac's notation):
\begin{equation}
 \begin{array}{r}
\langle \delta |\hat{H} | \gamma \rangle \equiv \\
\langle \delta |\hat{H}^{2} | \gamma \rangle \equiv 
      \end{array}
	\begin{array}{l}
(\langle \delta |\hat{H} \gamma \rangle +
\langle \gamma |\hat{H} \delta \rangle ^{*})/2 , \\
\langle \hat{H}\delta |\hat{H} \gamma \rangle .
       \end{array}
\label{ele}
\end{equation}
By using (\ref{int}), we obtain explicitly the following diagonal
matrix elements in the semiclassical limit $\hbar\rightarrow 0$,
\[ \begin{array}{r}
	i)  \\
	ii)  \\
	iii) 
   \end{array} \;
    \begin{array} {l}
         \langle \gamma | \gamma \rangle \rightarrow 1, \\
 \overline{E}_{\gamma}\equiv  \langle \gamma | \hat{H} | \gamma \rangle/
 \langle \gamma | \gamma \rangle \rightarrow E_{\gamma},  \\
 \sigma_{\gamma}^{2}\equiv\langle \gamma|\hat{H}^{2}|\gamma 
\rangle/ \langle \gamma | \gamma \rangle\! -\! \overline{E}_{\gamma}^{2}
 \rightarrow (\hbar\lambda)^{2}\; \overline{\dot{x}^2}/2  ,
	\end{array}
\]
with $ \overline{\dot{x}^2}=S(L)/mT\;$ the time average of 
$\dot{x}^2$ on the orbit.
Expression $iii)$ shows that the width $\sigma_{\gamma}$ of the resonance 
is asymptoticaly proportional to $\lambda$. Moreover, 
$\;\rho_{E} \;\sigma_{\gamma}\! =\! {\cal O}(\hbar^{-1})\;$ \cite{ac3}
shows that
a unique orbit cannot support a stationary state in the semiclassical
limit (to support an eigenfunction the width of a resonance needs to
satisfy 
$\rho_{E}\; \sigma < 1$). Of course, this result
is well known \cite{shn}.

We should say some words about symmetry.
If the system is time reversal it is easy to show that 
$\;\psi_{-\gamma}(x,y)=\psi_{\gamma}(x,y)^{*}\;$ and
$\;\hat{H}\psi_{-\gamma}(x,y)=(\hat{H}\psi_{\gamma}(x,y))^{*}\;$,
where $-\gamma$ is the time reversal partner of $\gamma$. If the system 
also includes a spatial symmetry $G$, it results that
$\;\psi_{G\gamma}(x,y)=G\psi_{\gamma}(x,y)\;$ and 
$\;\hat{H}\psi_{G\gamma}(x,y)=G\hat{H}\psi_{\gamma}(x,y)\;$. So,
to obtain real eigenfunctions inside a defined symmetry 
representation, we construct real resonances inside 
the same representation by using group theory \cite{lan}.

For low or medium energies we can evaluate  matrix elements  
directly on the domain; however, for high energies
or to obtain explicit expressions in terms of classical quantities
as $\hbar$ goes to zero, it is preferable 
to work on a surface of section \cite{bog2}. Let $\zeta$ be a differentiable 
curve with the coordinate $q$ along it, and $\eta$ perpendicular
to $\zeta$ at $q$ ($\eta\!=\!0$ on the curve). Suppose $\gamma$ crosses
the section at $q_{j}$ ($j\!=\!1,\ldots,m$) with angles
$\theta_{j}$; the corresponding positions on $\gamma$ being 
$x_{j}$ (see Fig.~\ref{fig}(a)). In a neighborhood of radio 
${\cal O}(\sqrt{\hbar})$ around the intersection point $j$, the
coordinates are related to order $\hbar$ by
$\;x-x_{j}=\sin(\theta_{j}) \;(q-q_{j})-\cos (\theta_{j})\;\eta\;$ 
and $\;y=\cos(\theta_{j}) \;(q-q_{j})+\sin (\theta_{j})\;\eta\;$.
Then, the restriction of $\psi_{\gamma}$ to $\zeta$, up to 
${\cal O}(\sqrt{\hbar})$, is a sum of
wave packets in one dimension with tangential momentum 
$p_{j}=m\;\dot{x}_{j}\sin (\theta_{j})$,
\begin{equation}
\varphi_{\gamma}(q)\equiv\psi_{\gamma}(x,y){\mid}_{\zeta}=\!
{\sum }_{j=1}^{m}\; \psi_{\gamma_{j}}(q) ,
\label{rest1}
\end{equation}
with $\;\psi_{\gamma_{j}}(q)= \exp[i\;p_{j}(q\!-\!q_{j})/\hbar]\;
\psi_{\gamma}[x_{j},\cos(\theta_{j})(q\!-\!q_{j})].$\\

If the system is bounded by a hard wall and we take this wall
like the Poincar\'e surface of section, $\psi_{\gamma}$ is 
null there up to order $\sqrt{\hbar}$ \cite{ac4}. In the
neighborhood of a bouncing point, $\psi_{\gamma}$
consists of two terms associated to the incoming and outgoing
trajectory such that the combination satisfies boundary conditions
\cite{bab} (see Fig.~\ref{fig}(b)). Then, working with its normal derivative
we define
\begin{equation}
\varphi_{\gamma}(q)\equiv - \frac{i \hbar}{2m}
\frac{\partial \psi_{\gamma}}{\partial \eta} (x,y){\mid}_{\zeta}=\!
{\sum }_{j} \dot{x}_{j} \cos (\theta_{j})
  \psi_{\gamma_{j}}(q).
\label{rest2}
\end{equation}
Moreover, $\hat{H} \varphi_{\gamma}(q) \equiv \hat{H}
\psi_{\gamma}(x,y){\mid}_{\zeta}$ for (\ref{rest1}) and 
$\hat{H} \varphi_{\gamma}(q)\equiv- (i \hbar/2m)
\partial \hat{H}\psi_{\gamma}/\partial \eta (x,y)
{\mid}_{\zeta}\;$ for (\ref{rest2})
are obtained to the leading
order from 
(\ref{rest1}) or (\ref{rest2}) taking into account that
\begin{equation}
\hat{H}\psi_{\gamma_{j}}(q)\simeq g_{\gamma}[x_{j},\cos(\theta_{j})
(q-q_{j})]\; \psi_{\gamma_{j}}(q) .
\label{hami}
\end{equation}
From now on $\varphi_{\gamma}(q)$ (Eq. (\ref{rest1}) or
(\ref{rest2})) is the object representing the resonance on the
section. In order to describe an effective Hilbert space 
we define a norm on $\zeta$, 
\begin{equation}
\langle \gamma | \gamma \rangle _{\zeta} \equiv \int_{\zeta}
\varphi_{\gamma}(q)^{*} \; \varphi_{\gamma}(q)\; f(q)\; dq ,
\label{norma}
\end{equation}
such that $i)$, $ii)$ and $iii)$ are satisfied.
Then, evaluating the leading term
of the integral in (\ref{norma}), it results in a classical criterium
for specifying $f(q)$
\begin{equation}
T={\sum }_{j=1}^{m}\;
f(q_{j})\; [ \dot{x}_{j}\;\cos (\theta_{j})]^{\mp 1} . 
\label{funcion}
\end{equation} 
The signs ($-$) and ($+$) correspond to (\ref{rest1}) 
and (\ref{rest2}) respectively.
We select a smooth real function $f(q)$, oscillating so slowly as 
possible, which
satisfies (\ref{funcion}) for all short {\bf P.O}s
required in the calculation. We notice that the existence of $f(q)$
is not guaranteed for all sections. A hint to choose a section could
be to make sure that the classical motion between consecutive points 
(the map) is simple.
For example, in billiards the motion between bounces with
the boundary is simple. On the other hand, 
eigenfunctions of billiards are reduced
to the boundary in terms of their normal derivatives, and the metric 
on the boundary 
is defined by setting $f(q)=\!2\;({\bf r.\hat{n}})_{(q)}/\dot{x}^{2}$
\cite{ver2}, with ${\bf \hat{n}}$ the unit outgoing normal to the boundary 
and ${\bf r}$ the position vector.
Then, condition (\ref{funcion}) (with sign ($+$)) reduces to 
\[
L={\sum }_{j=1}^{m}\; 2\; ({\bf r.\hat{n}})_{(q_{j})}
\cos (\theta_{j}) ,
\]
and this nice identity is valid for any {\bf P.O} in any
billiard; the demostration  being a simple geometrical problem.

Now we can evaluate matrix elements over the section
by using equations (\ref{rest1}) or (\ref{rest2}),
(\ref{hami}) and (\ref{norma}) [$\delta$ crosses the section at
$q_{k}$ ($k\!=\!1,\ldots,m'$) with angles $\theta_{k}$],
\begin{equation}
\langle \delta |\hat{O}  \gamma \rangle _{\zeta} \!=\!
\sum _{j,k=1}^{m,m'} \; A_{jk}\!
\int_{\zeta} \psi_{\delta_{k}}(q)^{*}
\hat{O} \psi_{\gamma_{j}}(q)\;f(q)\;dq.
\label{ele2}
\end{equation}
$A_{jk}\!=\!1$ for (\ref{rest1}) and 
\mbox{$A_{jk}=\dot{x}_{j}\dot{x}_{k} \cos(\theta_{j})\cos(\theta_{k})$} 
for (\ref{rest2}). Defining $z_{l}=-i \cos^{2}(\theta_{l}) \Gamma
(x_{l})/2$, $c_{l}=q_{l}+i p_{l}/2 z_{l}$ and 
$c_{jk}=(z_{j} c_{j}+z_{k}^* c_{k}^*)/(z_{j}+z_{k}^*)$, the Gaussian
integrals in (\ref{ele2}) are given by
\[
\frac{D_{jk}\; f(c_{jk})\; \exp[-B_{jk}/\hbar+i\;(\alpha_{j}-\alpha_{k})]}
{\sqrt{T_{\gamma}\;T_{\delta}\; \dot{x}_{j}\; \dot{x}_{k}\;
|Q(x_{j})\; Q(x_{k})| \;(z_{j}+z_{k}^*)/J}} .
\]
$B_{jk}\!=\!z_{j}z_{k}^*(c_{j}-c_{k}^*)^{2}/(z_{j}+z_{k}^*)+p_{j}^{2}/
4z_{j}+p_{k}^{2}/4z_{k}^*$ and $\alpha_{l}=S(x_{l})/\hbar-\phi(x_{l})/2$.
Moreover, $D_{jk}\!=\!1$ for $\hat{O}\!=\! \hat{1}$ and 
$D_{jk}=g_{\gamma}[x_{j}, \cos(\theta_{j})\sqrt{(c_{jk}-q_{j})^{2}+
\hbar/2(z_{j}+z_{k}^*)}]$ for $\hat{O}\!=\!\hat{H}$.
Then, Eq. (\ref{ele2}) gives
matrix elements in terms of classical quantities evaluated
at the intersection points of the orbits with the surface of section.

Finally, in order to obtain the eigenenergies and eigenfuctions of a
bounded chaotic Hamiltonian system in a given energy range, we procced
as follows. It is constructed the family of resonances of the 
shortest periodic
orbit $\gamma_{1}$, living in the required energy range. The density
of resonances associated to $\gamma_{1}$ is 
$\rho_{1}\simeq T_{1}/2 \pi \hbar$, with $T_{1}$ the period of $\gamma_{1}$.
Later, we do the same with the next shortest orbit $\gamma_{2}$, and so on
(using only primitive orbits).
The process stop when the whole density of resonances equals
the mean energy density $\rho_{E}$,
\begin{equation}
T_{H}\equiv 2 \pi \hbar \; \rho _{E} \simeq 2 \pi \hbar {\sum}_{k=1}^{N_{p.o}}
\rho_{k} \simeq {\sum}_{k=1}^{N_{p.o}} T_{k} .
\label{tiempo}
\end{equation}
Equation (\ref{tiempo}) is actually
impressive, it shows that the number of {\bf P.O}s $N_{p.o}$
required in the calculation is very little and increases at most
{\it linearly} with the Heisenberg time
$T_{H}$. More precisely 
$N_{p.o} \simeq h T_{H}/\ln (h T_{H})$, where $h$ is the
topological entropy.
For methods related to the trace formula 
$N_{p.o} \simeq \exp(h T_{H})/ (h T_{H})$.

Another interesting quantity is the number of resonances $N_{res}$
contributing to one eigenfunction.
This number is proportional to the mean dispersion (see $iii)$)
and to $\rho_{E}$
\begin{equation}
N_{res}\simeq 3.6 \hbar \rho_{E} \;\sqrt{< \lambda^{2}\;
\overline{ \dot{x}^{2}} >} ,
\label{dimension}
\end{equation}
with $<\;>$ the average over {\bf P.O}s (using the factor $3.6$
in (\ref{dimension}), the $99\%$ of an eigenfuntion is recovered). 
Then, we select $N$
resonances ($N\geq N_{res}$), consecutive in energy, and call them 
$\Gamma_{1},\ldots , \Gamma_{N}$. Later, by solving 
the following generalized eigenvalue problem
\begin{equation}
{\sum}_{j=1}^{N} \left(\langle \Gamma_{k} | \hat{H} | \Gamma_{j} \rangle -
 E \;\langle \Gamma_{k} | \Gamma_{j} \rangle \right) \xi_{j} =0\;,\;\;
\forall \;k,
\label{quantization}
\end{equation}
the eigenenergies $E$ and eigenvectors $\xi$ in the basis 
of resonances are given.

The main idea for the selection of resonances is to obtain a 
quasi-orthogonal basis of highly localized (in energy) functions. The
best way of satisfying quasi-orthogonality is to use short periodic
orbits. Now, for orbits with comparable periods we select the one
with minimum energy dispersion (see $iii)$). This analysis works for
hard chaos systems (where all {\bf P.O}s are unstable and isolated).
However, for systems with a fraction of regular motion, we need to
include the same fraction of regular functions in the basis. And, for
systems with a continuous family of neutral {\bf P.O}s (e.g. the
bouncing ball family in the stadium billiard), it is required
the corresponding fraction of phase space localized functions.

In conclusion we need to construct adequate functions in each
classically different region of phase space, the number of them
satisfying the required mean density (obtained semiclassically)
in each region. In a chaotic region, functions
(we call them resonances) are constructed with the shortest
{\bf P.O}s, and the number of them used to fill the Hilbert space
is very little as implied in Eq. (\ref{tiempo}). In order to obtain
the interaction between two given short {\bf P.O}s it is possible
to follow different strategies. Thinking at classical level it is
necessary to use at least a {\bf P.O} living in the neighborhood
of the previous ones. Then, to obtain all matrix elements, the period
of the orbits required in the full calculation would be of the order
of the Heisenberg time, and no advantage is reached with respect to
other approaches. However, in this article we showed that thinking
at quantum level, the interaction between short {\bf P.O}s can be
evaluated simply in terms of transversal excitations. In this way,
all the information required in the calculation is contained
in short {\bf P.O}s. 

We finish with some remarks.
i) This theory has been applied successfully to the desymmetrized
stadium billiard \cite{ver3}. ii) In this theory, eigenvalues
have an error ${\cal O}(\hbar)$ and eigenfunctions an error
${\cal O}(\sqrt{\hbar})$. iii)
We believe that the inclusion of transversal excitations in the
construction of resonances is the way of extending this theory to
higher orders in $\hbar$. iv)
The basis of resonances is particularly useful for parametric 
dependent systems. In fact, this is the {\it diabatic basis}, so 
difficult to find in chaotic systems \cite{ver}.
v) The evaluation of $\sigma_{\gamma}$ as a function of $n$
is a simple an efficient semiclassical measure of scars \cite{heller}.
vi) The phenomenon of chaos-assisted tunnelling can be studied without
external parameters \cite{boh}.

Supported in part by PICT97 03-00050-01015 and SECYT-ECOS. I 
would like to thank G. Carlo, M. Saraceno, R. Vallejos and D. Wisniacki for 
useful discussions.

\begin{figure}
\caption{ (a) Coordinates $(x,y)$ and $(q,\eta)$ defining the neighborhood
of the periodic orbit $\gamma$ and the section $\zeta$ respectively.
. (b) Idem (a) but the section
is given by a hard wall.}
\label{fig}
\end{figure}


\begin{references}
\bibitem{ac0} The monodromy matrix starting at $x\!=\!x_{0}$ is related
to $M(x)$ in a simple way: $M_{x_{0}}(L)\!=\!M(x_{0})M(L)M(x_{0})^{-1}$.
So, it is easy to find the required condition.
\bibitem{gausiana} The initial complex vector 
$\xi_{u}+\!i \;\xi_{s}$ defines at $x\!=\!0$ a transversal wave
packet with the following two properties: i) maximum overlap with
the wave packet obtained from it by a one period linealized evolution,
and ii) maximum concentration around the orbit minimizing non 
lineal contributions to the one period exact evolution.
\bibitem{ac2} An expression in that region
can be obtained by using the Airy function. There are no problems 
if the turning point occurs in a hard wall potential.
\bibitem{ac3} The energy density for two degrees of fredom
is ${\cal O}(\hbar^{-2})$.
\bibitem{ac4} For billiards, the natural Poincar\'e surface of section
is the boundary where $\psi_{\gamma}$ is null (for Dirichlet conditions).
\end{references}
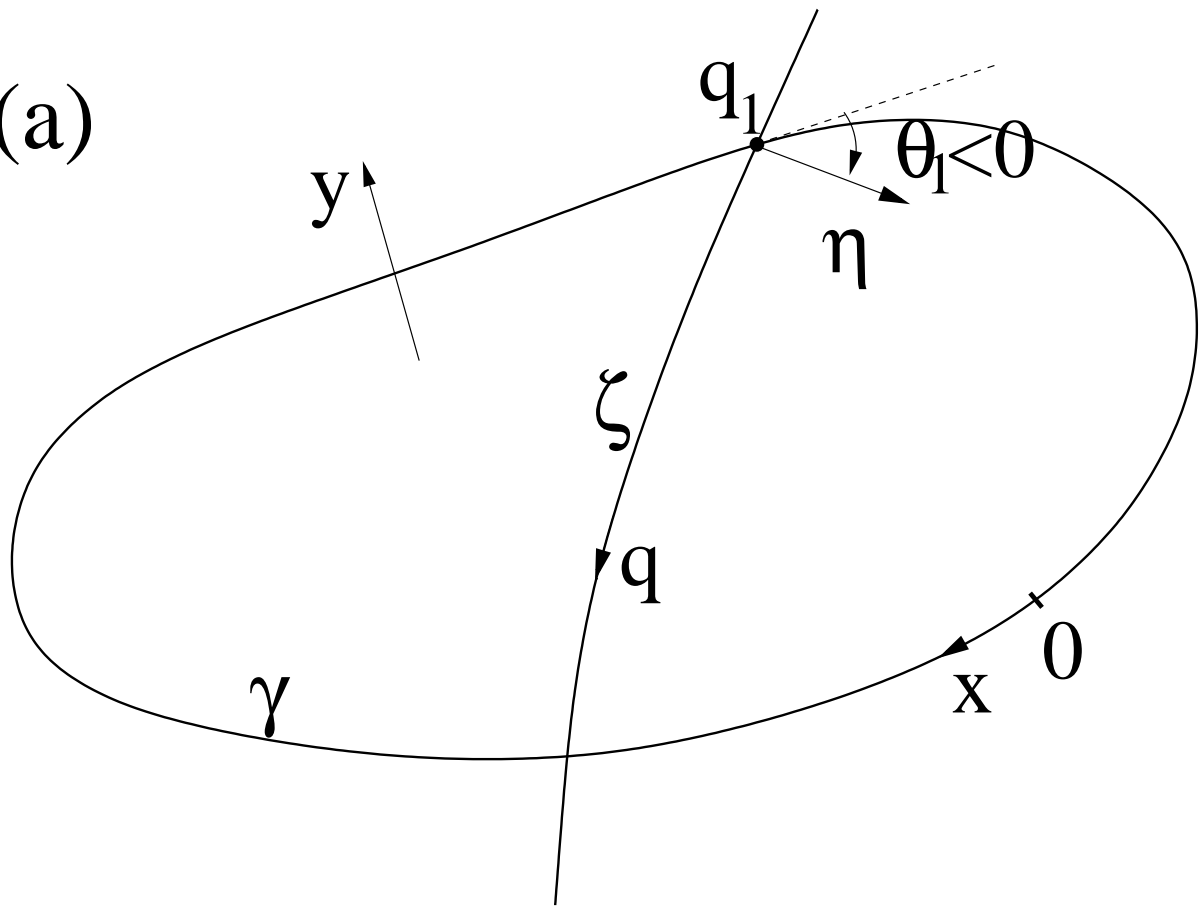
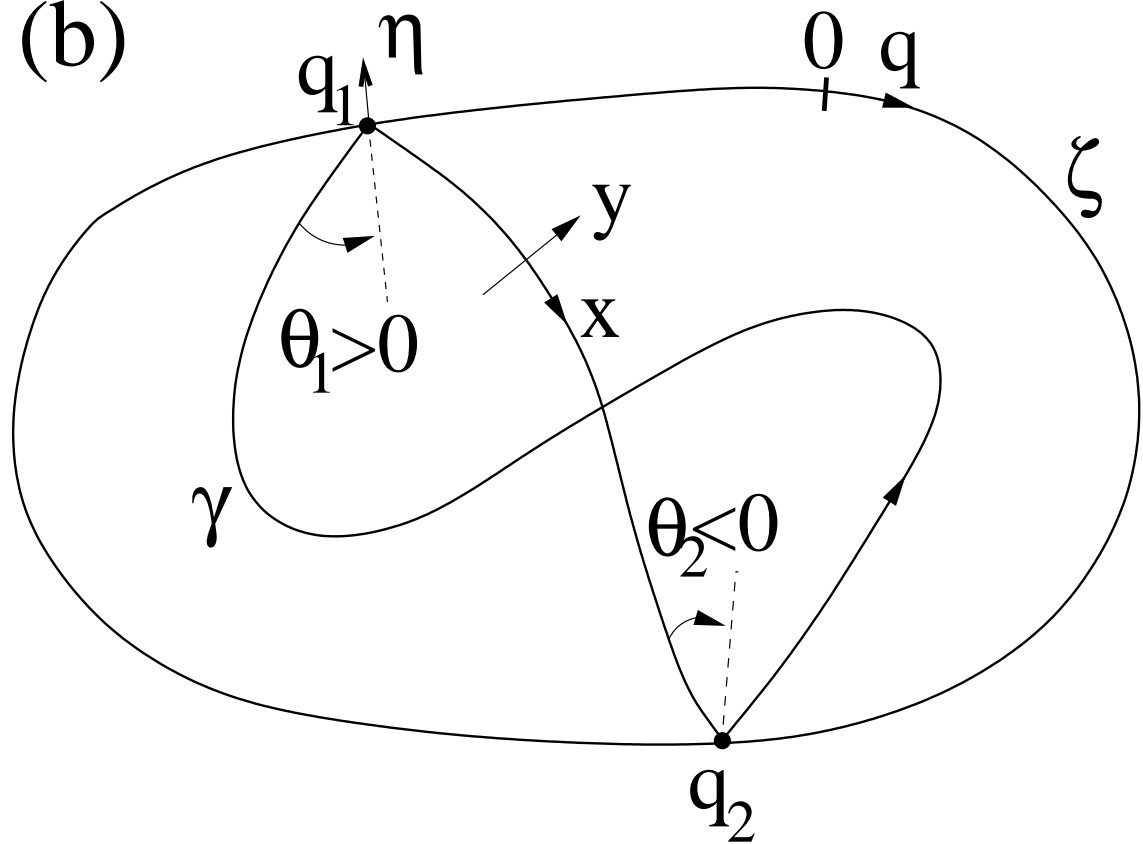
\end{document}